\documentstyle[prl,aps,amsfonts,amssymb,epsfig]{revtex}
\begin{document}
\draft 
\twocolumn[\hsize\textwidth\columnwidth\hsize\csname 
@twocolumnfalse\endcsname
\underline{Physical Review Letters, in press.}

\title{Direct two-magnon optical absorption in $\alpha^{\prime}$-NaV$_2$O$_5$ }
\author{A. Damascelli, D. van der Marel, M. Gr\"uninger, 
C. Presura, and T.T.M. Palstra}
\address{Material Science Center, University
 of Groningen, Nijenborgh 4, 9747 AG Groningen, The Netherlands}
\author{J. Jegoudez, and A. Revcolevschi} 
\address{Laboratoire de Chimie des Solides, Universit$\acute{e}$ de 
Paris-sud, 
B$\hat{a}$timent 414, F-91405 Orsay, France}
\date{10 March 1998}
\maketitle
\begin{abstract}
We investigated the temperature-dependent optical conductivity of 
$\alpha^{\prime}$-NaV$_2$O$_5$ in the energy range 4 meV-4 eV. 
The intensities and the polarization dependence of the detected electronic excitations 
give a direct indication for a broken-parity electronic 
ground-state and for a non-centrosymmetric 
crystal structure of the system in the high-temperature phase. A direct
two-magnon optical absorption process, proposed in this Letter, is in 
quantitative agreement with the optical data. By analyzing the optically 
allowed phonons at various temperatures 
above and below the phase transition, we  conclude that a second-order 
change to a larger unit cell takes place below 34 K. 
\end{abstract}
%
%\pacs{PACS numbers: ????}
\vskip2pc]
\narrowtext
CuGeO$_3$ and $\alpha^{\prime}$-NaV$_2$O$_5$ are both insulating
compounds exhibiting a temperature dependence of the magnetic susceptibility 
which has been discussed in terms of the one-dimensional (1D) S=1/2-Heisenberg 
spin chain. Moreover, a phase transition has been observed, at 
14 K for CuGeO$_3$\cite{hase} and at 34 K for 
$\alpha^{\prime}$-NaV$_2$O$_5$\cite{isobe}, which is characterized by the opening of a 
spin gap and
by superlattice reflections\cite{boucher,fujii}. Together these observations have been 
interpreted as indications of a spin-Peierls (SP) phase transition. For
both compounds, and in particular for CuGeO$_3$, the 2D character can not be
neglected and is the subject of intensive investigations.
In this Letter we demonstrate that, in contrast with the situation in
conventional SP systems, the low-energy spin excitations 
of $\alpha^{\prime}$-NaV$_2$O$_5$ carry a finite electric dipole moment. We
probe this dipole moment directly in our optical experiments, and
use it to measure the spin-correlations as a function of temperature.
\\
The basic building blocks of the crystal structure of 
$\alpha^{\prime}$-NaV$_2$O$_5$ are linear chains of alternating
V and O ions, oriented along the {\em b} axis\cite{carpy}. These chains are 
grouped into sets of two, forming a ladder, with the rungs oriented along the
{\em a} axis (see Fig.~1). The rungs are formed by two V ions, one on each leg of the 
ladder, bridged 
by an O ion. The V-O distances along the rungs are shorter than along 
the legs, implying a stronger bonding along the rung. In the {\em a-b} plane 
the ladders are shifted half a period along the {\em b} axis relative to  
their neighbors.
\\
The average charge of the V ions of +4.5 implies an
occupation of the V 3d band of half an electron which, in 
principle, implies metallic behavior. Lattice deformations can however 
lift the degeneracy between the two V sites. The early X-ray 
diffraction (XRD) analysis of Carpy and Galy\cite{carpy} indicated 
the non-centrosymmetric space group {\em P2$_1$mn}. In this 
structure it is possible to identify well-distinct magnetic 
V$^{4+}$ (S=1/2) and non-magnetic V$^{5+}$ (S=0) chains running 
along the {\em b} axis of the crystal and alternating
each other along the {\em a} axis. This configuration 
would  be responsible for the 1D character of the high temperature 
susceptibility\cite{mila} and for the SP transition, possibly involving 
dimerization within the V$^{4+}$ chains\cite{isobe}. The 
insulating character of $\alpha^{\prime}$-NaV$_2$O$_5$ would be an 
obvious consequence of  having a 1D 1/2-filled Mott-Hubbard system. 
Recently, this structural analysis has been questioned  and the 
centrosymmetric space group {\em Pmmn} was proposed\cite{meetsma,smolinski}. 
In this context, it is still possible to recover 
an insulating ground state assuming that the {\em d} electron, supplied by 
the two V ions forming a rung of the ladder, is not attached to a particular 
V site but is shared in a V-O-V molecular bonding orbital
along the rung\cite{smolinski}. However, the debate on the appropriate space group and  
electronic ground-state for  
$\alpha^{\prime}$-NaV$_2$O$_5$ is still open. While single-crystal XRD refinements
indicate an inversion center, small deviations in atomic positions up to 0.03 \AA\ could 
not be excluded\cite{meetsma}.
\\
In this Letter we present an investigation of the temperature-dependent 
optical conductivity of pure high-quality single crystals of 
$\alpha^{\prime}$-NaV$_2$O$_5$. Based on our analysis
of the charge-transfer (CT) absorption edge at

\noindent
\begin{figure}[htb]
\centerline{\epsfig{figure=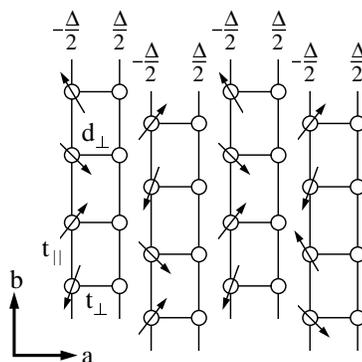,width=5cm,clip=}}
 \caption{Sketch of the ladder structure in the {\em a-b} 
plane of $\alpha^{\prime}$-NaV$_2$O$_5$. The circles represent the 
V sites having on-site energies $\pm \Delta/2$ on adjacent legs, alternatively. 
The arrows represent  the electrons 
which are mainly distributed over legs characterized by lower potential 
energy for the V sites.
}
\label{fig1}
\end{figure}

\noindent
1 eV we provide strong support
for a broken-parity ground-state of the quarter-filled ladder. This lack
of inversion 
symmetry provides a mechanism for two-magnon optical 
absorption, and is shown to be in agreement with the observed oscillator
strength and polarization dependence of the experimental data. Further 
experimental support for these 'charged magnons' is provided 
by the strong line-shape anomalies of the optical phonons.
\\
High-quality single crystals of $\alpha^{\prime}$-NaV$_2$O$_5$
were grown by high temperature solution growth from a vanadate mixture flux. 
The crystals with dimensions of  
$\sim 1\times$3$\times$0.3 mm$^3$ were 
aligned by Laue diffraction and mounted in a liquid He flow 
cryostat to study the temperature dependence of the normal incidence 
reflectivity between 30$\,$ and $\,$30\,000 cm$^{-1}$ in the
temperature range 4-300 K. The optical conductivity
was calculated from the reflectivity using Kramers-Kronig relations.
\\
The {\em a} and {\em b}-axis reflectivity spectra at 
T=4 K and T=40 K are displayed in Fig.~2 up to 1000 cm$^{-1}$, which 
covers the full phonon spectrum. In Fig.~3 we present the optical
conductivity. The {\em b}-axis optical conductivity (along the chains) 
has no electronic absorption in the far infrared, which is characteristic 
of an insulating material. Along the {\em a} direction (along the rungs) we 
observe a broad band of weak optical absorption in the far
and mid-infrared range. For both directions we observe a strong absorption
at 8000 cm$^{-1}$ (1 eV), in agreement with Ref.~9. We observe 6 
{\em a}-axis  phonon modes at 
90 (weak), 137, 256, 518, 740 (weak) and 938 cm$^{-1}$ (weak), and 4 
{\em b}-axis  modes at 177, 225 (weak), 371 and 586 cm$^{-1}$. 
The factor group analysis gives 7 (15) {\em a}-axis and 4 (7) {\em b}-axis 
infrared active phonons for the {\em Pmmn} ({\em P2$_1$mn}) space group. 
As some of the modes may have escaped  detection (due to a small oscillator strength, 
to masking by the electronic infrared continuum or because the frequency is too low 
for our setup), neither of the two space groups can be ruled out on the basis of the 
number of experimentally observed  phonons.
\\
Let us now analyze the absorption band at 1 eV. An obvious candidate for such a
strong optical transition is the on-rung CT involving the 
two V sites. We model the {\em j}-th rung with the Hamiltonian:
 \parbox{7.6cm}{\begin{eqnarray*}
         H_j &=& 
        t_{\bot} \sum_{\sigma}\left\{L_{j\sigma}^{\dagger}R_{j\sigma}
         + R_{j\sigma}^{\dagger}L_{j\sigma}\right\}
         + \Delta n_{C}  \\
         &+& U \left\{n_{jR\uparrow}n_{jR\downarrow}
         + n_{jL\uparrow}n_{jL\downarrow}\right\} ,
   \end{eqnarray*}} \hfill
   \parbox{.8cm}{\begin{eqnarray}\end{eqnarray}} 

\noindent
where $L_{j\sigma}^{\dagger}$ ($R_{j\sigma}^{\dagger}$) creates an electron 
with spin $\sigma$ on the left-hand (right-hand) V site of the {\em j}-th 
rung, $U$ is the on-site Hubbard repulsion, $t_{\bot}$ is the on-rung hopping 
parameter, $n_{C}\!=\!\sum_{j}(n_{jR}\!-\!n_{jL})/2$ is the charge displacement 
operator, and $\Delta$ is the potential energy difference between the two 
sites. 
For the symmetric ladder $\Delta = 0$. For a single electron
per rung, the two solutions are lob-sided bonding and anti-bonding combinations  
$|\tilde{L}\!>=\!u|L\!>\!+v|R\!>$, and $|\tilde{R}\!>\!=u|R\!>\!-v|L\!>$

\noindent
\begin{figure}[t]
\centerline{\epsfig{figure=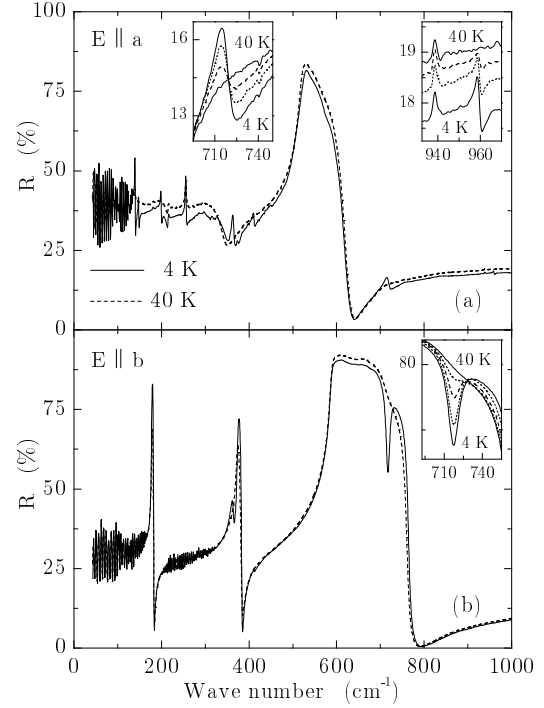,width=7cm,clip=}}
 \caption{Reflectivity spectra of $\alpha^{\prime}$-NaV$_2$O$_5$ for 
 $\vec{E}\!\parallel\!\vec{a}$ (a) and $\vec{E}\!\parallel\!\vec{b}$ (b)
 below (4 K) and above (40 K) the phase transition. Insets: detailed temperature 
 dependence of  some of the new phonon lines detected for T$<$T$_{\rm{SP}}$.
}
\label{fig2}
\end{figure}

\noindent
where $1/(2uv)^2\!=\!1\!+\!(\Delta/2t_{\bot})^2$. The splitting 
between these two eigenstates, $E_{CT}\!=\!\sqrt{\Delta^2+4t_{\bot}^2}$,  
corresponds to 
the photon {\em energy} of the optical absorption (1 eV). 
\\
A second crucial piece of information is provided by the {\em intensity} 
of the absorption. We use the following expression, which is exact for 
one electron on two coupled tight-binding orbitals\cite{sum}:
 \begin{equation}
 \int_{peak} \sigma_{1}(\omega) d\omega = 
 \pi e^2 N d_{\bot}^2 t_{\bot}^2 \hbar^{-2} E_{CT}^{-1} ,
 \end{equation}
where {\em N} is the volume density of the rungs and $d_{\bot}$=3.44  \AA\ is the 
distance between the two V ions on the same rung (see Fig.~1). This way we calculated from the
spectra  $|t_{\bot}|\!\approx 0.3$ eV. Combining this number with
$E_{CT}\!=\!1$ eV we obtain  
$\Delta \!\approx\! 0.8$ eV. The corresponding two eigenstates have
90 \% and 10 \% character on either side of the rung. Therefore the
valence of the two V-ions is 4.1 and 4.9, respectively. The optical 
transition at 1 eV is essentially a CT excitation 
from the occupied V 3d state at one side of the rung to the empty
3d state at the opposite side of the same rung.
\\
Let us now turn to the infrared continuum for 
$\vec{E}\!\parallel\!\vec{a}$ [an 
enlarged view is
 given in inset (a) of  Fig.~3]. As the
range of frequencies coincides with the low energy-scale spin excitations,
the most likely candidates for this continuum are excitations involving
two spin flips.  Another reason for this assignment is the opening, for 
T$<$T$_{\rm{SP}}$\cite{unpu}, of a gap in the optical conductivity of 
17$\pm$3 meV ({\em i.e.}, approximately twice the spin gap 
value\cite{fujii,weiden}). The presence of 
{\em two} V states ($|L\!>$ and 
$|R\!>$) per spin, along with the broken left-right parity of the ground state, 
gives rise to a fascinating behavior of the spin flips: Consider a 
small fragment of the ladder with only two rungs, with one spin
per rung. Each rung is described by the Hamiltonian defined above. 
In the ground state of this cluster each spin resides in a
$|L\!>$ orbital, with some admixture of $|R\!>$. Let us now include
the coupling of the two rungs along the legs, using:
 \begin{equation}
   t_{\parallel}
  \sum_{\sigma}\left\{R^{\dagger}_{1,\sigma}R_{2,\sigma} 
  + L^{\dagger}_{1,\sigma}L_{2,\sigma} + H.C. \right\} .
 \end{equation}
If the two spins are parallel, the inclusion of this term has
no effect, due to the Pauli-principle. If they form a $S=0$ state, 
the ground state can gain some kinetic energy along the legs by
mixing in states where two electrons reside on the same rung. Working
in the limit that $U\!\rightarrow\!\infty$, these states 
have one electron in the $|L\!>$ and one in the $|R\!>$ state on 
the same rung. As a result, there is a net dipole displacement of the
singlet state compared to the triplet state: 
{\em The spin-flip excitations carry a finite electric dipole moment.}
Using a perturbation expansion in $t_{\parallel}/E_{CT}$, and working
in the limit $U\!\rightarrow\!\infty$, the exchange
coupling constant between two spins on neighboring rungs becomes:
 \begin{equation}
 J_{\parallel}=\frac{8t_{\parallel}^2 t_{\perp}^2} {[\Delta^2+4t_{\perp}^2]^{3/2}} .
 \end{equation}
The coupling to infrared light with  $\vec{E}\!\parallel\!\vec{a}$ 
can  now be included   using the dipole approximation. The
only effect is to change the potential energy of the $|R\!>$ states relative
to the $|L\!>$ states. In other words, we have to replace $\Delta$ with
$\Delta\!+\!q_e d_{\bot} E_a$, where $q_e$ is the electron charge, and 
$E_a$ is the component of the electric
field along the rung. By expanding $J_{\parallel}$ we obtain the 
spin-photon coupling:  
 \begin{equation}
  H_S=q_m d_{\bot} E_a h_{S}=q_m d_{\bot} E_a \sum_{j}\vec{S}_j\cdot\vec{S}_{j + 1} ,
 \end{equation}
where $q_m\!=\!q_e \frac{3J_{\parallel}\Delta}{\Delta^2+4t_{\perp}^2}$
is the effective charge involved in a double
spin-flip transition. For a symmetrical ladder, where $\Delta\!=\!0$, the
effective charge vanishes, and the charged magnon effect disappears.
In that case higher-order processes like the phonon assisted 
spin-excitations, 
 considered by Lorenzana and Sawatzky\cite{lorenzana}, can still contribute. These
are probably responsible for the weak mid-infrared continuum in the
$\vec{E}\parallel \vec{b}$ spectra. 
\\
Taking the values of $t_{\bot}$ and $\Delta$ calculated 
from the optical data and $|t_{\parallel}|\!\approx\!0.2$ eV\cite{harrison}  
we obtain $J_{\parallel}\!\approx\!30$ meV (comparable to the reported values ranging from
38\cite{weiden} to 48 meV\cite{isobe}), 
and $q_m/q_e\!=\!0.07$. Using the  functions: 

\noindent
\parbox{7.4cm}{\begin{eqnarray*}
  g_{S}(T)&\equiv& 4<<h_{S}^2>-<h_{S}>^2>_{T} ,  \\
  g_{C}(T)&\equiv& 4<<n_{C}^2>-<n_{C}>^2>_{T} ,  
  \end{eqnarray*}} \hfill
   \parbox{.8cm}{\begin{eqnarray}\end{eqnarray}}

\noindent
\begin{figure}[htb]
\centerline{\epsfig{figure=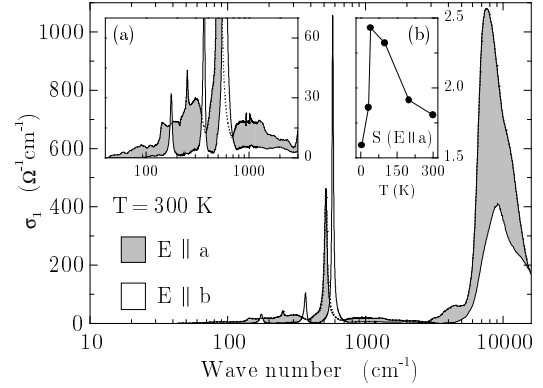,width=7cm,clip=}}
 \caption{Optical conductivity of $\alpha^{\prime}$-NaV$_2$O$_5$ at 300 K for 
 $\vec{E}\!\parallel\!\vec{a}$ and $\vec{E}\!\parallel\!\vec{b}$. Inset (a): enlarged 
 view of $\sigma_{1}(\omega)$ from 40 to 3000 cm$^{-1}$. Inset (b): oscillator 
 strength of the electronic continuum for $\vec{E}\!\parallel\!\vec{a}$ plotted 
 versus temperature.
}
\label{fig3}
\end{figure}

\noindent
we can express the intensity of the spin-fluctuations
 relative to the
CT excitations in terms of the effective charge: 
 \begin{equation}
 \frac{\int_{S}\sigma_{1}(\omega)d\omega}{\int_{CT}\sigma_{1}(\omega)d\omega}
 = \frac{q_m^2g_{S} E_{S}}
        {q_e^2g_{C} E_{CT}} .
 \label{relint}
 \end{equation}
It is easy to prove that $g_{C}\!=\!(2uv)^2$. For the
present parameters,  $g_{C}\!\simeq\!0.4$. $E_{S}$ is the 
average energy of the infrared spin fluctuation spectrum and is of the 
order of 0.1 eV. The spin-correlation function $g_{S}$ depends on the 
details of the many-body wave function of the spins. For
an antiferromagnetic chain $g_{S}\!=\!1$, whereas for
a random orientation of the spins it is zero. In between it depends on
the probability of having fragments of three  
neighboring spins ordered antiferromagnetically.
Hence the maximum relative intensity in Eq. (\ref{relint}) is $\sim$0.0014. The 
experimental value is $\sim$0.0008 at T=300 K, in good agreement 
with the numerical 
estimate. Note, that a finite value of $g_{S}$ requires either
terms in the Hamiltonian which do not commute with $h_{S}$, or an
antiferromagnetic broken symmetry of the ground state.
\\
Remarkable is also the temperature dependence of the spin-fluctuation
continuum observed for $\vec{E}\!\parallel\!\vec{a}$. The oscillator strength, obtained by 
integrating $\sigma_{1}(\omega)$ (with phonons subtracted) up to 800 cm$^{-1}$, is displayed 
in inset (b) of  Fig.~3. It increases upon cooling down the sample from room 
temperature to T$_{\rm{SP}}$ and rapidly decreases for T$<$T$_{\rm{SP}}$.
Within the discussion given above, this increase marks an increase in short
range antiferromagnetic correlations of the chains. Below the phase transition, 
nearest neighbor spin-singlet correlations become dominant, 
and  $g_{S}$ is suppressed. 
\\
Upon cooling down the sample below T$_{\rm{SP}}$=34 K, significant 
changes occur in the phonon spectrum (see Fig.~2). 
Contrary to  CuGeO$_3$\cite{sp}, where 
we observed in reflectivity a single, very weak additional phonon,  10 new 
lines are  detected for $\vec{E}\!\parallel\!\vec{a}$ 
($\omega_{\rm{TO}}$\,$\approx$\,101, 127, 147, 199, 362, 374, 410, 
450, 717, and 960  cm$^{-1}$), and 7 for $\vec{E}\!\parallel\!\vec{b}$ 
($\omega_{\rm{TO}}$\,$\approx$\,102, 128, 234, 
362, 410, 718, and 960 cm$^{-1}$), some of them

\begin{figure}[b]
\centerline{\epsfig{figure=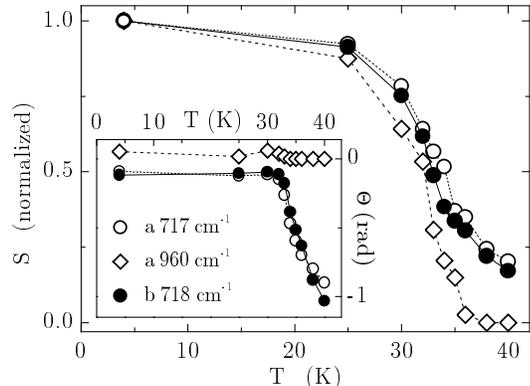,width=7cm,clip=}}
 \caption{Temperature dependence of the normalized oscillator 
 strength of the zone boundary  modes observed  at 717 
  and 960 cm$^{-1}$,
for $\vec{E}\!\parallel\!\vec{a}$, and  
at 718 cm$^{-1}$, for $\vec{E}\!\parallel\!\vec{b}$. 
Inset:  Fano asymmetry parameter for these lines.
}
\label{fig4}
\end{figure}

\noindent
  showing
 a quite large  oscillator strength. An enlarged view of few of these 
phonons is given in the insets of Fig.~2.
The observation of almost the same frequencies for both axes, which is an important information 
for a full understanding of the  structural distortion, is an intrinsic 
property  of the low temperature phase. Experimental errors, like 
polarization leakage or sample misalignment, are excluded from the well-defined 
anisotropic phonon spectrum of  the undistorted phase.  
The larger intensity of the lines, with respect to the case of 
CuGeO$_3$\cite{sp}, reflects the larger lattice distortions in 
$\alpha^{\prime}$-NaV$_2$O$_5$ on going through the phase transition. 
At the same time the strong line-shape anomalies of all infrared
active phonons for $\vec{E}\!\parallel\!\vec{a}$ indicate a strong coupling
  to the spin-fluctuation continuum. We
like to speculate here that this is due to coupling of the
 phonons to the electric dipole moment of the charged magnons.
\\ 
We carefully investigated the activated modes at $\approx$718 and 960 cm$^{-1}$, 
 at temperatures ranging from 4 to 40 K. We fitted  these phonons to a Fano profile
because the 718 cm$^{-1}$ modes show an asymmetrical line-shape. The results of the 
normalized oscillator strength {\em  S} are plotted versus temperature in Fig.~4. 
In the inset we also present the asymmetry parameter $\Theta$, defined in 
such a way that for $\Theta$=0 a Lorentz line-shape is 
recovered\cite{fesi}. {\em S} has a similar behavior for the
three different lines. However, the 960 cm$^{-1}$ peak vanishes at 
T$_{\rm{SP}}$ whereas the  two 718 cm$^{-1}$ modes have still a finite intensity at 
T=40 K and, as a matter of fact,  disappear only  
for T$>$60-70 K. At the same time,   the line-shape of the 960 
cm$^{-1}$ mode is perfectly lorentzian at all temperatures, whereas
the two other  phonons show a consistently increasing asymmetry 
for T$>$32 K. From these results we  conclude that the second-order 
character of the phase transition is nicely shown by the behavior of {\em  S} for the 960
cm$^{-1}$ folded mode. On the other hand, pretransitional fluctuations
manifest themselves in the finite intensity of the 718 cm$^{-1}$ modes above T$_{\rm{SP}}$. 
Similar pretransitional fluctuations have been observed below 70 K in the course of a study 
of the propagation of ultrasonic waves along the chain direction of 
$\alpha^{\prime}$-NaV$_2$O$_5$\cite{fertey}.

In conclusion, by a detailed analysis of the optical conductivity we 
provide  direct evidence for the broken-parity electronic 
ground-state and  for the non-centrosymmetric crystal structure 
of the high temperature phase of  $\alpha^{\prime}$-NaV$_2$O$_5$.
 We show that a direct  two-magnon optical absorption process 
 is responsible for the low frequency continuum observed 
 perpendicularly to the chains. By analyzing the optically 
 allowed phonons at various temperatures, we conclude that a second-order 
 change to a larger unit cell takes place below 34 K, with a 
 fluctuation regime  extending over a very broad temperature range.
 
We gratefully acknowledge A. Lande, M. Mostovoy, D.I. Khomskii, and G.A. Sawatzky  
for stimulating discussions. We thank D. Smirnov, J. Leotin,  M. Fischer 
and P.H.M. van Loosdrecht for many useful comments, and C. Bos, 
A. Meetsma  and J.L. de Boer for assistance. This investigation was supported 
by the Netherlands Foundation for Fundamental Research on Matter 
(FOM) with financial aid from the Nederlandse Organisatie voor 
Wetenschappelijk Onderzoek (NWO).

\end{document}